\documentclass[10pt]{article}
\usepackage{amsmath,amssymb,cite}

\newtheorem{theorem}{Theorem}[section]   

\newtheorem{lemma}[theorem]{Lemma}
\newtheorem{proposition}[theorem]{Proposition}
\newtheorem{corollary}[theorem]{Corollary}

\newtheorem{remark}[theorem]{Remark}
\numberwithin{equation}{section}
\numberwithin{theorem}{section}

\newcommand{\due}[2]{\genfrac{}{}{0pt}{1}{#1}{#2}}  
\newcommand{\mc}[1]{{\mathcal #1}}
\newcommand{\bb}[1]{{\mathbb #1}}
\newcommand{\proof}{\noindent{\bf Proof.} }
\newcommand{\qed}{\hfill $\square$\medskip}

\newcommand{\address}[1]{\begin{itemize}\item[]\rm\raggedright #1 \end{itemize}}\def\eads#1{\address{E-mail: #1}}\def\mailto#1{{\tt #1}}\def\ams#1{\begin{itemize}\item[]\rm AMS classification scheme numbers: #1\par\end{itemize}} \newcommand{\rme}{\mathrm{e}}\newcommand{\rmi}{\mathrm{i}}\newcommand{\rmd}{\mathrm{d}}

\begin{document}

\title{Navier-Stokes equations on the flat cylinder with vorticity production on the boundary}

\author{C. Boldrighini$^1$ and P. Butt\`a$^1$}

\maketitle

\address{$^1$ Dipartimento di Matematica, SAPIENZA Universit\`a di Roma, P.le Aldo Moro 2, 00185 Roma, Italy}
\eads{\mailto{boldrigh@mat.uniroma1.it}, \mailto{butta@mat.uniroma1.it}}  
\ams{35Q30, 76D05, 76N10}

\begin{abstract}
We study the two-dimensional Navier-Stokes system on a flat cylinder $\mc C$ with the usual Dirichlet boundary conditions for the velocity field $u$, $u|_{\partial \mc C} = 0$.  We formulate the problem as an infinite system of ODE's for the natural Fourier components of the vorticity, and the boundary conditions are taken into account by adding a vorticity production at the boundary.  We prove equivalence to the original Navier-Stokes system and show that the decay of the Fourier modes is exponential for any positive time in the periodic direction, but it is only power-like in the other direction.
\end{abstract}

\section{Introduction}
\label{sec:1}

We study the two-dimensional incompressible Navier-Stokes (NS) equations on the flat cylinder $\mc C := \bb T \times [0,\pi]$, where $\bb T$ is the one-dimensional torus obtained by identifying the end-points of the interval $[-\pi,\pi]$. We consider the usual boundary value problem in   absence of external forces,  
\begin{equation}
\label{p1}
\left\{\begin{array}{l} \partial_t u + (u\cdot\nabla) u = \Delta u -\nabla p, \\ \nabla\cdot u = 0, \\ u|_{t=0} = u^{(0)}, \end{array} \right.
\end{equation}
with Dirichlet boundary conditions, 
\begin{equation}
\label{p2}
u\big|_{\partial \mc C} = 0.
\end{equation}
In \eqref{p1}, $u=(u_1,u_2)$ is the velocity field, $p$ is the pressure, and the viscosity is taken equal to $1$. We denote by $x=(x_1,x_2)\in (-\pi,\pi]\times [0,\pi]$ the coordinates on $\mc C$ and set $\nabla^\perp := (-\partial_{x_2},\partial_{x_1})$.  The vorticity is defined as
\begin{equation}
\label{p3}
\omega := \nabla^\perp \cdot u = \partial_{x_1}u_2-\partial_{x_2}u_1.
\end{equation}
If $u\in C^1(\mc C;\bb R^2)$ is a solenoidal vector field  (i.e.~$\nabla\cdot u = 0$) satisfying the boundary condition \eqref{p2} then, as shown in Lemma~\ref{lem:1} in the following section, the vorticity has zero mean on $\mc C$, and $u$ is uniquely determined by $\omega$ and can be represented as
\begin{equation}
\label{p4}
u = \nabla^\perp \Delta_N^{-1}\omega,
\end{equation}
where $\Delta_N$ is the Laplacian on $\mc C$ with zero Neumann boundary conditions. Therefore, an evolution equation for $\omega$, which is equivalent, assuming sufficient smoothness, to the NS system, can be derived by taking the curl of both sides of \eqref{p1}-\eqref{p2} and using \eqref{p4},
\begin{equation}
\label{p5}
\left\{\begin{array}{l} \partial_t\omega + u\cdot\nabla\omega = \Delta\omega, \\ \partial_{x_1}\Delta_N^{-1}\omega|_{\partial \mc C} = 0, \\ \omega|_{t=0} = \nabla^\perp\cdot u^{(0)}. \end{array} \right.
\end{equation}
The new formulation has the advantage of   eliminating the pressure, but, on the other hand, the Dirichlet boundary conditions for $u$ are replaced by  the  linear non-local condition \eqref{p5}$_2$  for the Laplacian operator appearing in \eqref{p5}$_1$, which is involved and difficult to   handle. 

In this paper, following an approach which was introduced in the physical literature \cite{B}, we  take into account the condition \eqref{p5}$_2$  by adding a suitable vorticity production term on the boundary.  A rigorous proof that the resulting integro-differential problem  is equivalent  to the usual NS system was given by Benfatto and Pulvirenti for the half-plane \cite{BP}. To our knowledge, there is up to now no general result in this sense.   

More precisely, we treat the Laplacian in \eqref{p5}$_1$ as the operator with Neumann boundary conditions, which preserves vorticity, and add on the right side a vorticity production term  in order to satisfy the condition \eqref{p5}$_2$. In this way we obtain the equation,
\begin{equation}
\label{p7}
\left\{\begin{array}{l} \partial_t\omega + u\cdot\nabla\omega = \Delta_N\omega + f\delta_{\partial\mc C}, \\ \partial_{x_1}\Delta_N^{-1}\omega|_{\partial \mc C}  = 0, \\ \omega|_{t=0} = \nabla^\perp\cdot u^{(0)}. \end{array} \right.
\end{equation}
As the boundary is made of two separate pieces, we have   
\begin{equation}
\label{p8}
f\delta_{\partial\mc C}(x,t) = f_1(x_1,t)\delta(x_2) + f_2(x_1,t)\delta(x_2-\pi).
\end{equation}
The functions $f_1(x_1,t)$, $f_2(x_1,t)$ have to be determined in such a way that \eqref{p7}$_2$ is satisfied. They can be uniquely determined under  the conditions
\begin{equation}
\label{p9}
\int_{\bb T}\!\rmd x_1\, f_j(x_1,t) =0, \qquad j=1,2.
\end{equation}
As we shall see, the conditions \eqref{p9} are needed if we want that the dynamics defined by  \eqref{p7} preserves the average value of the vorticity, i.e.,   
\begin{equation}
\label{p9a}
\frac{\rmd}{\rmd t} \int_{\mc C}\!\rmd x\, \omega(x,t) = 0,
\end{equation}
and gives the usual balance equation for the component along the periodic direction  of the total momentum of the fluid in absence of external forces,
\begin{equation}
\label{p9b}
\frac{\rmd}{\rmd t} \int_{\mc C}\!\rmd x\, u_1(x,t) = \int_{\bb T}\!\rmd x_1\, \big[\omega(x_1,0,t) - \omega(x_1,\pi,t)\big].
\end{equation}
By  \eqref{p9a},  we have $\int_{\mc C}\rmd x\, \omega(x,t) = 0$ for all $t>0$ if  $\int_{\mc C}\rmd x\, \omega(x,0) = 0$ at the initial time $t=0$,  so that  $u(x,t) = \nabla^\perp \Delta_N^{-1}\omega(x,t)$ makes sense for all $t>0$. 

We will give a precise sense to \eqref{p7} as a set of ordinary differential equations (ODE's) for the Fourier modes in the basis of the eigenfunctions of the Laplacian with Neumann boundary conditions, and will prove equivalence to the NS system \eqref{p1}-\eqref{p2}. 

Our methods are inspired by the recent works \cite{DDS1,DDS2}, which show how to  obtain, by mainly elementary methods,  deep results on the regularity of the solutions to the NS system, in particular on the decay of the Fourier modes. Such results are a natural continuation of the well-known works on the NS system on the flat two-dimensional torus \cite{G,MS,FT,T1}.
 
Our main result is that, under some mild assumption on the initial data, which imply continuity of the vorticity, we prove that for any positive $t>0$ the natural Fourier modes decay exponentially fast in  the periodic direction, but only as an inverse square in the other direction. This result should be compared with the case of the two-dimensional flat torus, where the decay is exponential in both directions.

In addition to the strong regularization of the Fourier modes in the periodic direction we also have a  mild regularization in the other direction as well. Namely, for any $t>0$,  $u$ is continuous together with its first and second derivatives,  up to the boundary of $\mc C$. As a consequence, if the initial data are regular enough our solutions are classical solutions in the sense of Ladyzhenskaya \cite{L}. 
 
A similar picture for the decay of the Fourier modes  was first shown to hold for a plane NS problem  with different boundary conditions \cite{DDS1}. Our results corroborate the opinion of the authors of that paper that such a picture holds in general  for the NS system in a bounded plane region with smooth boundary.

\section {Notation and formulation of the main results}
\label{sec:2}

We begin the section with a justification of the representation \eqref{p4}.
\begin{lemma}
\label{lem:1}
If $u\in C^1(\mc C;\bb R^2)$ is solenoidal, i.e.~$\nabla\cdot u = 0$, and satisfies the Dirichlet boundary conditions \eqref{p2} then $\int_{\mc C}\rmd x\, \omega(x) = 0$, where $\omega$ is the vorticity defined in \eqref{p3}, and $u$ can be represented as  $u = \nabla^\perp \psi$, where the ``stream function'' $\psi\in  C^2(\mc C;\bb R)$ is the unique (up to a constant) solution of the Poisson problem on $\mc C$ with zero Neumann conditions,
\begin{equation}
\label{pp1} 
\Delta \psi = \omega, \qquad \partial_{x_2} \psi\big|_{\partial \mc C} =0. 
\end{equation}  
Moreover $\int_0^\pi\rmd x_2\, u_1(x_1, x_2)= c$ is constant in $x_1$, and $\psi$ is constant on the two pieces of the boundary, $\psi(x_1,0) = c_0$, $\psi(x_1,\pi) = c_1$, with the constraint
\begin{equation*}
\psi(x_1, 0) - \psi(x_1, \pi) = c_0-c_1 = \int_0^\pi\!\rmd x_2\,  u_1(x_1,x_2) = c.
\end{equation*}
\end{lemma}
\proof The condition $\int_{\mc C}\rmd x\, \omega(x) = 0$ follows immediately by periodicity in $x_1$ and the boundary condition $u_1\vert_{\partial \mc C}=0 $. Moreover, by periodicity in $x_1$, solenoidality, and the boundary condition $u_2\vert_{\partial \mc C}=0 $ it is easily seen that the integrals
\begin{equation*}
I_1(x_1) = \int_0^\pi\!\rmd x_2\, u_1(x_1, x_2), \qquad  I_2(x_2) = \int_{\bb T}\!\rmd x_1\,  u_2(x_1, x_2)
\end{equation*}
are constant, i.e.~$I_1'(x_1) \equiv 0$,  $I_2'(x_2) \equiv 0$, and in fact $I_2(x_2) \equiv 0$.  

As $\nabla \cdot u =0$, we can write $u = \nabla^\perp \psi$, and the condition $I_2(x_2)=0$ implies that $\int_{\bb T}\rmd x_1\,\partial_{x_1} \psi(x_1,x_2)=0$, so that $\psi$ is periodic in $x_1$. Moreover $\nabla^\perp \cdot \nabla^\perp \psi = \Delta \psi$ and $\psi$ is necessarily a solution of the boundary value problem \eqref{pp1}, which  has a unique solution, up to a constant.  

The other boundary condition $u_2|_{\partial \mc C} =0$, which is not guaranteed by the problem \eqref{pp1},  implies that $\psi(x_1, \pi)$ and $\psi(x_1, 0)$ are constants, and their difference is clearly equal to $I_1(x_1) = c$.
\qed

We translate system \eqref{p7} into an infinite system of ODE's for the components of the vorticity with respect to the orthogonal basis of eigenfunctions of the Laplacian $\Delta_N$. A function $\phi\in L_2(\mc C)$  can be expanded in such basis as
\begin{equation*}
\phi(x) = \sum_{k_1\in\bb Z} \phi_{k_1,0} \, \rme^{\rmi k_1 x_1} + 2 \sum_{k_1\in\bb Z}\sum_{k_2\ge 1} \phi_{k_1,k_2}\, \rme^{\rmi k_1 x_1} \cos (k_2x_2),
\end{equation*}
where the coefficients of the generalized Fourier series are given by
\begin{equation*}
\phi_{k_1,k_2} = \frac{1}{2\pi^2} \int_{\mc C}\!\rmd x\, \phi(x)\, \rme^{-\rmi k_1 x_1} \cos(k_2x_2).
\end{equation*}
For computations it is often convenient to write the series in  a different way.    Extending $\phi_{k_1,k_2}$ by parity to negative $k_2$ by setting  $\hat\phi_{k_1,k_2} = \phi_{k_1,|k_2|}$ for any $k=(k_1,k_2)\in\bb Z^2$, we obtain
\begin{equation}
\label{p10}
\phi(x) = \sum_{k\in\bb Z^2} \hat \phi_{k_1,k_2} \, \rme^{\rmi k\cdot x}.
\end{equation}

We shall   assume that all  Fourier expansions  converge fast enough, so that all formal multiplications and  term-by-term differentiations which we need are legitimate. Such  assumptions are later  proved to be true. 

We first compute the  Fourier coefficients  of the transport term $u\cdot\nabla\omega$.  Given $\omega$ with $\omega_{0,0}=0$, if $\omega_{k_1,k_2}$ are the   coefficients of $\omega$ in the expansion \eqref{p10},  the stream function $\psi$ has   coefficients $\psi_{k_1,k_2} = - |k|^{-2} \omega_{k_1,k_2}$. Setting $k^\perp = (-k_2,k_1)$ and recalling $u = \nabla^\perp \psi$, we then find
\begin{equation}
\label{nn}
u(x) = -\rmi \sum_{\due{k\in\bb Z^2}{k\ne (0,0)}}  k^\perp \, \frac{\hat\omega_{k_1,k_2}}{k^2} \, \rme^{\rmi k\cdot x},\quad \nabla\omega(x) = \rmi \sum_{k\in \bb Z^2}  k\, \hat\omega_{k_1,k_2}\, \rme^{\rmi k\cdot x},
\end{equation}
\begin{equation}
\label{agg1}
[u\cdot \nabla \omega](x) = \sum_{k\in\bb Z^2} \hat R_{k_1,k_2} \, \rme^{\rmi k\cdot x}, \qquad \hat R_{k_1,k_2} = \sum_{\due{j+\ell = k}{j\ne (0,0)}} \frac{j^\perp\cdot \ell}{j^2} \, \hat\omega_{j_1,j_2}\,\hat\omega_{\ell_1,\ell_2}.
\end{equation}
Observing that $\hat R_{k_1,k_2} $ is odd in $k_2$, i.e.,   $\hat R_{k_1,k_2} = - \hat R_{k_1,-k_2}$, we find
\begin{equation*}
[u\cdot \nabla \omega](x) = 2\rmi \sum_{k_1\in\bb Z}\sum_{k_2\ge 1} \hat R_{k_1,k_2}\, \rme^{\rmi k_1 x_1} \sin (k_2x_2).
\end{equation*}
In order to go back the Neumann basis we use the expansion  \begin{equation*}
\sin(qy) = \sum_{p\in\bb Z} A_{q,p}\,\cos(py), \qquad A_{q,p} =
\frac{\delta_\mathrm{odd}(q+p)}\pi \frac{2q}{q^2-p^2}, 
\end{equation*}  
where $\delta_\mathrm{odd}(n)=1$ [resp.~$\delta_\mathrm{odd}(n)=0$] is $n$ is odd [resp.~even]. Therefore,
\begin{equation*}
[u\cdot \nabla \omega](x) = \sum_{k\in\bb Z} \hat N_{k_1,k_2} \, \rme^{\rmi k_1 x_1} \cos (k_2x_2), \qquad \hat N_{k_1,k_2} = \rmi\sum_{h_2\in\bb Z} \hat R_{k_1,h_2}\, A_{h_2,k_2}.
\end{equation*}
Clearly $\hat N_{k_1,-k_2} = \hat N_{k_1,k_2}$, so that, setting  $N_{k_1,k_2} = \hat N_{k_1,k_2}$ for $k_2\ge 0$, we find
\begin{equation}\label{agg3}
[u\cdot \nabla \omega](x) = \sum_{k_1\in\bb Z} N_{k_1,0} \, \rme^{\rmi k_1 x_1} + 2 \sum_{k_1\in\bb Z}\sum_{k_2\ge 1} N_{k_1,k_2}\, \rme^{\rmi k_1 x_1} \cos (k_2x_2).
\end{equation}
Taking into account \eqref{agg1}, the coefficients of the expansion \eqref{agg3} are written as 
\begin{equation}
\label{p13}
N_{k_1,k_2}[\omega] = \rmi\sum_{h_2\in\bb Z} \frac{\delta_\mathrm{odd}(h_2+k_2)}\pi \frac{2h_2}{h_2^2-k_2^2} \,\sum_{\due{j+\ell = (k_1,h_2)}{j\ne (0,0)}} \frac{j^\perp\cdot \ell}{j^2} \, \hat\omega_{j_1,j_2}\,\hat\omega_{\ell_1,\ell_2}.
\end{equation}
In terms of the Fourier components the boundary condition \eqref{p7}$_2$ reads
\begin{equation*}
\rmi \sum_{\due{k\in\bb Z^2}{k\ne (0,0)}}  k_1 \, \frac{\hat\omega_{k_1,k_2}}{k^2} \,  \rme^{\rmi k_1 x_1} \cos(k_2x_2)\; \Bigg|_{\partial\mc C} = 0, 
\end{equation*}
and splits into two  equations for the two components ($x_2=0$ and $x_2=\pi$) of $\partial \mc C$, which hold for   any $x_1\in (-\pi,\pi]$ if and only if 
\begin{equation*}
\sum_{k_2\in\bb Z} \frac{\hat\omega_{k_1,k_2}}{k^2} = 0, \qquad \sum_{k_2\in\bb Z} (-1)^{k_2} \, \frac{\hat\omega_{k_1,k_2}}{k^2} \qquad \forall\, k_1\ne 0.
\end{equation*}
Adding and subtracting and going back to the   components $\omega_{k_1,k_2}$ we finally get,
\begin{equation}
\label{p14}
\sum_{k_2,+} \frac{\omega_{k_1,k_2}}{k^2} = 0, \qquad \sum_{k_2,-} \frac{\omega_{k_1,k_2}}{k^2} = 0 \qquad \forall\, k_1\ne 0,
\end{equation}
where we use for brevity the notation 
\begin{equation*}
\sum_{s,+} a_s = a_0 + 2\sum_{i\ge 1} a_{2i}, \qquad \sum_{s,-} a_s = 2 \sum_{i\ge 1} a_{2i-1}.
\end{equation*}
 
Observe that, as  $\omega|_{t=0} = \nabla^\perp\cdot u^{(0)}$ and $u^{(0)}$ satisfies the Dirichlet boundary conditions, the initial data are ``well prepared'', i.e,  $\omega_{0,0}=0$ and \eqref{p14} are verified at $t=0$.

We can now write the NS equations as an infinite set of coupled  ODE's for the components $\omega_{k_1, k_2}(t)$, $k_1\in \bb Z$, $k_2\ge 0$.
Expanding both sides of \eqref{p7}$_1$ we obtain  
\begin{equation}
\label{p11}
\dot\omega_{k_1,k_2}(t) + N_{k_1,k_2}[\omega(t)] = - k^2 \omega_{k_1,k_2}(t) + f_{\pm,k_1}(t),\quad k_1\in\bb Z,\; k_2\ge 0,
\end{equation}
where the $+$ [resp.~$-$] sign is chosen for $k_2$ even [resp.~odd], and
\begin{equation}
\label{p12}
f_{\pm,k_1}(t) = f_{1,k_1}(t) \pm f_{2,k_1}(t), \qquad f_{j,k_1}(t) =
\frac{1}{2\pi^2}\int_{\bb T}\!\rmd x_1\, f_j(x_1,t)\, \rme^{\rmi k_1 x_1}.
\end{equation}
Equations \eqref{p11} are completed by the expression \eqref{p13} for the quadratic term $N_{k_1, k_2}$, and conditions \eqref{p14}. 

Moreover, for $k_1=0$ we need the conditions \eqref{p9} which now read $f_{\pm,0}(t)=0$. These conditions follow, as we claimed in the introduction, and will now show, from conditions  \eqref{p9a} and \eqref{p9b}. In fact, for \eqref{p9a}, observe that if $\omega$ satisfies \eqref{p14} then 
\begin{equation}
\label{N00}
N_{0,0}[\omega] = 2\rmi \sum_{j_1\in\bb Z} j_1 \sum_{j_2,\ell_2\in\bb Z} \frac{\delta_\mathrm{odd}(j_2+\ell_2)}\pi \, \hat\omega_{-j_1,\ell_2}\,\frac{\hat\omega_{j_1,j_2}}{j^2} = 0,
\end{equation}
so that $\dot\omega_{0,0}(t) = 0$ implies $f_{+,0}(t)=0$. As $\omega_{0,0}(0)=0$, this choice also gives $\omega_{0,0}(t) \equiv 0$. For the second one, we have
\begin{eqnarray*}
&& \frac{\rmd}{\rmd t} \int_{\mc C}\!\rmd x\, u_1(x,t) = - \frac{\rmd}{\rmd t} \int_{\mc C}\!\rmd x\, \partial_{x_2} \Delta_N^{-1}\omega (x,t)  
 = - \frac{\rmd}{\rmd t} \sum_{k_2\in \bb Z} 4\pi\delta_{\rm odd}(k_2)\frac{\hat\omega_{0,k_2}(t)}{k_2^2} \\ && ~~~~~~~~ = - 4 \pi\sum_{k_2,-} \frac{\dot\omega_{0,k_2}(t)}{k_2^2}  = 4\pi \sum_{k_2,-} \left\{ \omega_{0,k_2}(t)  + \frac{N_{0,k_2}[\omega(t)] - f_{-,0}(t)}{k_2^2} \right\} \\ && ~~~~~~~~ = \int_{\bb T}\!\rmd x_1\, \big[\omega(x_1,0,t) - \omega(x_1,\pi,t)\big] + 4\pi \sum_{k_2,-}\frac{N_{0,k_2}[\omega(t)] - f_{-,0}(t)}{k_2^2}.
\end{eqnarray*}
The claim is then a straightforward consequence of the following lemma. 

\begin{lemma}
\label{lem:2}
For any $\omega$ satisfying $\omega_{0,0}=0$ and \eqref{p14},
\begin{equation*}
\sum_{k_2,-} \frac{N_{0,k_2}[\omega]}{k_2^2} = 0.
\end{equation*}
\end{lemma}
\proof By the expression  \eqref{p13}, as $j^\perp\cdot \ell=j_1h_2$ for $j+\ell = (0,h_2)$, we have
\begin{eqnarray*}
\sum_{k_2,-} \frac{N_{0,k_2}[\omega]}{k_2^2} & = & \frac{2\rmi}{\pi} \sum_{k_2,-} \frac{1}{k_2^2} \sum_{h_2, +} \frac{h_2^2}{h_2^2-k_2^2} \sum_{\due{j+\ell = (0,h_2)}{j\ne (0,0)}} \frac{j_1}{j^2} \, \hat\omega_{j_1,j_2}\,\hat\omega_{\ell_1,\ell_2} \\ & = & \frac{2\rmi}{\pi} \sum_{h_2, +}\; \sum_{\due{j+\ell = (0,h_2)}{j\ne (0,0)}} \frac{j_1}{j^2} \, \hat\omega_{j_1,j_2}\,\hat\omega_{\ell_1,\ell_2} \sum_{k_2,-} \frac{h_2^2}{k_2^2(h_2^2-k_2^2)} \\ & = & 2\pi\rmi\sum_{n\ge 1}\; \sum_{\due{j+\ell = (0,2n)}{j\ne (0,0)}} \frac{j_1}{j^2} \, \hat\omega_{j_1,j_2}\,\hat\omega_{\ell_1,\ell_2},
\end{eqnarray*}
where in the last line we used the equalities
$ \frac{h_2^2}{k_2^2(h_2^2-k_2^2)} =   \frac{1}{k_2^2} -   \frac{1}{k_2^2 - h_2^2}$ and 
\begin{equation*}
\sum_{k_2,-} \frac{1}{k_2^2 - (2n)^2} = \left\{ \begin{array}{ll} {\displaystyle \frac{\pi^2}2} & \text{if $n=0$,}\\ \\ {\displaystyle \frac{\pi}{4 n}\tan(\pi n) = 0} & \text{if $n\ge 1$.}  \end{array}\right.
\end{equation*}
We thus obtain
\begin{equation*}        
\sum_{k_2,-} \frac{N_{0,k_2}[\omega]}{k_2^2} = 2\pi\rmi\sum_{j_1\ne 0}\left\{ \sum_{\ell_2,+} \omega_{-j_1,\ell_2} \sum_{j_2,+} \frac{\omega_{j_1,j_2}}{j^2} + \sum_{\ell_2,-} \omega_{-j_1,\ell_2} \sum_{j_2,-} \frac{\omega_{j_1,j_2}}{j^2} \right\} \; j_1,
\end{equation*}
which vanishes if $\omega$ satisfies \eqref{p14}. The lemma is proved. 
\qed 

As we will show in the next section, the conditions \eqref{p14} uniquely  determine the functions $f_{\pm,k_1}(t)$, for $k_1\neq 0$, as a quadratic integral functional of $\omega_{k_1, k_2}(s)$, $k_2\geq 0$, $s\in [0, t]$, so that the infinite system of coupled ODE's \eqref{p11} is in fact a system of integro-differential equations for the functions   
\begin{equation*}
\{\omega_{k_1,k_2}(t);\,  k_1\in\bb Z,\, k_2\ge 0\},
\end{equation*}  
satisfying an initial condition such that $\omega_{0,0}(0)=0$ and conditions \eqref{p14} are verified at $t=0$. We recall that such conditions imply $\omega_{0,0}(t)=0$ for all $t\ge 0$.

The infinite discrete problem defined in this way is the main object of our paper. In particular we will prove  that, under certain conditions on the initial data, it is equivalent to the NS problem \eqref{p1}-\eqref{p2}.

In what follows, by ``weak solution" of the NS problem \eqref{p1}-\eqref{p2} we mean the standard definition such as in  \cite{T1}. Our main result is the following theorem. 
\begin{theorem}
\label{thm:2}
Let $\omega_{k_1,k_2}(0)$ satisfy $\omega_{0,0}(0)=0$, \eqref{p14}, and the inequalities
\begin{equation}
\label{idi}
\left|\omega_{k_1,k_2}(0)\right| \le \frac{D_0}{|k|^{\alpha}\big(1+|k_1|^\beta\big)} \qquad \forall\, k_1\in\bb Z,\, k_2\ge 0,\, k\ne (0,0),
\end{equation}
with $1 < \alpha < 2$, $\beta\ge 0$, and $D_0>0$. Then there exist real numbers $D_1,\nu>0$ (depending on $D_0,\alpha,\beta$), and a unique solution $\{\omega_{k_1,k_2}(t);\,  k_1\in\bb Z,\, k_2\ge 0\}$ to equations \eqref{p11} and \eqref{p14} which satisfies for all $t\ge 0$ the inequalities
\begin{equation}
\label{tde}
\left|\omega_{k_1,k_2}(t)\right| \le \frac{D_1\, \rme^{-\nu(1+|k_1|)t}}{|k|^{\alpha}\big(1+|k_1|^\beta\big)} \qquad \forall\, k_1\in\bb Z,\, k_2\ge 0,\, k\ne (0,0).
\end{equation}
Moreover, for each $t_0>0 $ there is a constant $\tilde D_1=\tilde D_1(D_0,t_0,\alpha,\beta)$ such that
\begin{equation}
\label{idib}
|\omega_{k_1,k_2}(t)| \le \frac{\tilde D_1\, \rme^{-\nu(1+|k_1|)t/2}}{k^2} \qquad \forall\, t\ge t_0.
\end{equation}
Finally, the velocity field $u(x,t) := \nabla^\perp\Delta_N^{-1}\omega(x,t)$ associated to the vorticity
\begin{equation}
\label{o=}
\omega(x,t) = \sum_{k\in\bb Z^2}\hat \omega_{k_1,k_2}(t) \, \rme^{\rmi k\cdot x},
\end{equation}
is a weak solution to the NS system \eqref{p1}-\eqref{p2}. 
\end{theorem}
The decay estimate \eqref{tde} guarantees continuity  of $\omega(x,t)$ and  $C^\infty$ regularity  with respect to the periodic variable $x_1$ for any $t>0$, up to the border $\partial \mc C$.  The stronger estimate \eqref{idib} only implies that $\partial_{x_2}\omega(\cdot,t)$ is in $L^2(\mc C)$ for any $t>0$. In fact, the vorticity possesses  higher regularity, as stated by the following corollary.
\begin{corollary}
\label{cor:1}
For each $t>0$ the velocity field  $u(x,t) := \nabla^\perp\Delta_N^{-1}\omega(x,t)$  is  continuous and twice differentiable in $x_1, x_2$ up to the boundary  $\partial \mc C$. 
\end{corollary}
We remark that if $\alpha> \frac 32$ and $\alpha+\beta>2$ then $u^{(0)}\in W^{2,2}(\mc C)$, whence $u(x,t)$ is a classical solution of the NS system \eqref{p1}-\eqref{p2}, see e.g.~\cite[Chap.~6, Thm.~7]{L}. 

The proof of  Theorem~\ref{thm:2} and Corollary~\ref{cor:1} is deferred to Section  4. The following Section 3 is devoted to  an existence and uniqueness theorem for local solutions, with global solutions for small initial data, which is a first step in the proof of  Theorem~\ref{thm:2}.

\section{Local solutions}
\label{sec:3}

We first formulate a local existence theorem, which also provides global solutions for small initial data. 
\begin{theorem}
\label{thm:3}
Let $\omega_{k_1,k_2}(0)$, with  $\omega_{0,0}(0)=0$, satisfy \eqref{p14}, and the inequalities \eqref{idi} for some  $1<\alpha<2$, $\beta\ge 0$, and $D_0>0$. Then there exist a time $T_0=T_0(D_0,\alpha,\beta)$ and a constant $D_2=D_2(D_0,\alpha,\beta)$ such that there is a unique solution of equations \eqref{p11} and \eqref{p14} for $t\in[0,T_0]$,  which satisfies    the inequalities
\begin{equation}
\label{tde1}
\left|\omega_{k_1,k_2}(t)\right| \le \frac{D_2\, \rme^{-(1+|k_1|)t/4}}{|k|^{\alpha}\big(1+|k_1|^\beta\big)} \qquad \forall\, k_1\in\bb Z,\, k_2\ge 0,\, k\ne (0,0).
\end{equation}
Moreover, if $D_0$ is sufficiently small,   the corresponding solution is global and the estimate \eqref{tde1} is valid for any $t\ge 0$.
\end{theorem}

The proof of  Theorem~\ref{thm:3} is based on some constructions and  preliminary results. We begin by establishing a convenient representation of the components of the boundary term $f_{\pm, k_1}(t)$, $k_1\in \bb Z$,  in terms of the vorticity components $\omega_{k_1, k_2}(s)$, $k_2\geq 0$, $0\leq s \leq t$. 
To this end, by Duhamel's formula, we rewrite \eqref{p11} into the integral equation,
\begin{equation}
\label{p16}
\omega_{k_1,k_2}(t) = \rme^{-k^2t}\omega_{k_1,k_2}(0) + \int_0^t\!\rmd s\, \rme^{-k^2(t-s)} \, \Big\{f_{\pm,k_1}(s) - N_{k_1,k_2}[\omega(s)]\Big\}.
\end{equation}
The constraints \eqref{p14} give rise to a Volterra integral equation of the first kind for $f_{\pm,k_1}(t)$,
\begin{equation}
\label{p18}
\sum_{k_2,\pm} \frac{1}{k^2}  \int_0^t\!\rmd s\, \rme^{-k^2(t-s)} \, f_{\pm,k_1}(s) = g_{\pm,k_1}[t;\omega],
\end{equation}
where
\begin{equation}
\label{p17}
g_{\pm,k_1}[t;\omega] = \sum_{k_2,\pm} \frac{1}{k^2} \left\{ - \rme^{-k^2 t} \omega_{k_1,k_2}(0) + \int_0^t\!\rmd s\, \rme^{-k^2(t-s)} \, N_{k_1,k_2}[\omega(s)]\right\}.
\end{equation}
For any $\alpha,\beta\ge 0$ and $T>0$ we introduce the Banach space $\Omega_{\alpha,\beta,T}$ consisting of continuous functions $\omega(t) = 
\{\omega_{k_1,k_2}(t);\, k_1\in\bb Z,\, k_2\ge 0\}$, $t\in [0,T]$,  with $\omega_{0,0}(t) = 0$, endowed with the norm $\|\omega\|_{\alpha,\beta,T}$ where, for any $t\in [0,T]$,
\begin{equation}
\label{norm}
\|\omega\|_{\alpha,\beta,t} := \sup_{s\in [0,t]}\, \sup_{k_1\in\bb Z} \, \sup_{k_2\ge 0}\,\left|\omega_{k_1,k_2}(s)\right| \, \rme^{(1+|k_1|)s/4}\, 
|k|^{\alpha}\,\left(1+|k_1|^\beta\right).
\end{equation}  
The proof of  Theorem~\ref{thm:3} will be obtained by a contraction argument in the space $\Omega_{\alpha,\beta,T}$. We begin with some preliminary lemmata.
\begin{lemma}
\label{lem:3}
The Volterra equation of the first kind for the unknown function $a(t)$,
\begin{equation}
\label{p18bis}
\sum_{k_2,\pm} \frac{1}{k^2}  \int_0^t\!\rmd s\, \rme^{-k^2(t-s)} \, a(s) = b(t), \qquad k_1\ne 0,
\end{equation}
where  $b(t)$ is a bounded differentiable function with $b(0)=0$,  has a unique solution which can be represented as 
\begin{equation}
\label{A=}
a(t) = \int_0^t\!\rmd s\, G^\pm_{k_1} (t-s)\, b'(s) + \int_0^t\!\rmd s\, H^\pm_{k_1} (t-s)\, b(s). 
\end{equation}
Here, denoting by $\Gamma(\cdot)$ the Euler Gamma function,   $G^\pm_{k_1}$ is given by
\begin{equation}
\label{G=}
G^\pm_{k_1} (t) := \frac 2\pi d_\pm(k_1) \left[\delta(t) + \frac{\rme^{-k_1^2t}}{\sqrt t} \sum_{n=1}^4 \frac{d_\pm(k_1)^n}{\Gamma(n/2)}\,t^{(n-1)/2}\right],
\end{equation}
\begin{equation}
\label{d=}
d_\pm(k_1) := k_1\left[\tanh\left(\frac\pi 2 k_1\right)\right]^{\pm 1},
\end{equation}
and $H^\pm_{k_1} (t)$ is a continuous function such that, for each $0<\gamma <1$, 
\begin{equation}
\label{H}
H^\pm_{k_1} (t) \le B_\gamma \, |k_1|^3 \,\exp\left[-(1-\gamma)\, k_1^2t\right],
\end{equation}
with $B_\gamma$ a positive constant.
\end{lemma}
\proof Denoting by $\tilde F(\lambda) := \int_0^\infty\!\rmd t\,\rme^{-\lambda t}\, F(t)$, $\lambda\in \bb C$, the Laplace transform of the function $F(t)$, equation \eqref{p18bis} becomes
\begin{equation}
\label{le1}
\left(\sum_{k_2,\pm} \frac{1}{k^2} - \sum_{k_2,\pm} \frac{1}{k^2+\lambda} \right)\tilde a(\lambda) = \lambda\, \tilde b(\lambda),
\end{equation}
where $\Re\lambda> -k_1^2$. By the well known expansions,
\begin{equation*}
\tan z = - \sum_{n=1}^\infty \frac{2z}{z^2-(2n-1)^2\pi^2/4}, \qquad
\cot z = \frac 1z + \sum_{n=1}^\infty \frac{2z}{z^2-n^2\pi^2},
\end{equation*}
we have 
\begin{equation}
\label{Q=}
\sum_{k_2,\pm} \frac{1}{k^2+\lambda}  = \frac \pi 2 \, \frac{\phi_\pm(\sqrt{k_1^2+\lambda})}{\sqrt{k_1^2+\lambda}}, \qquad \phi_\pm(z) := \left[\tanh\left(\frac{\pi}2 z \right)\right]^{\mp 1}.
\end{equation}
In particular, by  \eqref{d=}, we find
\begin{equation}
\label{d==}
\sum_{k_2,\pm} \frac{1}{k^2} = \frac \pi 2 \, \frac{\phi_\pm(|k_1|)}{|k_1|} = \frac \pi 2 \,\frac 1{d_\pm(k_1)},
\end{equation}
so that the equation  \eqref{le1} reads 
\begin{equation}
\label{Atilde=1}
\tilde a(\lambda) = \frac 2\pi \frac{d_\pm(k_1)\,\lambda \, \sqrt{k_1^2+\lambda}}{\sqrt{k_1^2+\lambda}- d_\pm(k_1)\, \phi_\pm(\sqrt{k_1^2+\lambda})} \, \tilde b(\lambda).
\end{equation}
The right side of \eqref{Atilde=1} is the Laplace transform of the convolution of $b(t)$ with a kernel which is  the sum of a $\delta$-function, a singular   integrable kernel, and a regular term.  In fact, by the definition of the Euler Gamma function, we have
\begin{equation*}
\frac{\Gamma(\alpha+1)}{z^{\alpha+1}} = \int_0^\infty\!\rmd t\, \rme^{-z t} \, t^\alpha \qquad (\alpha>-1),
\end{equation*}
so that the Laplace transform of the kernel $G^\pm_{k_1}(t)$ defined in \eqref{G=} is  
\begin{equation*}
\tilde G^\pm_{k_1}(\lambda) = \frac 2\pi d_\pm(k_1)\, \sum_{n=0}^4 \left[\frac{d_\pm(k_1)}{\sqrt{k_1^2+\lambda}}\right]^n.
\end{equation*}
Since $b(0)=0$ implies $\lambda\, \tilde b(\lambda) = \widetilde{b'}(\lambda)$,   equation  \eqref{Atilde=1} becomes
\begin{equation}
\label{Atilde=2}
\tilde a(\lambda) = \tilde G^\pm_{k_1}(\lambda) \widetilde{b'}(\lambda) +  \tilde H^\pm_{k_1}(\lambda) \tilde b(\lambda),
\end{equation}
where
\begin{eqnarray}
\label{Htilde}
\nonumber
&& \!\!\!\!\!\!\!\!\!\!\!\!\!\!\!\tilde H^\pm_{k_1}(\lambda) \; =  \; \frac 2\pi d_\pm(k_1)\, \lambda \sum_{n=0}^4 \left[\frac{d_\pm(k_1)}{\sqrt{k_1^2+\lambda}}\right]^n\left[\phi_\pm\left(\sqrt{k_1^2+\lambda}\right) - 1 \right] \\ && ~~~~~~ + \, \frac 2\pi d_\pm(k_1)\,\left[\frac{d_\pm(k_1)}{\sqrt{k_1^2+\lambda}}\right]^4 
\frac{\lambda}{\sqrt{k_1^2+\lambda} - d_\pm(k_1)\, \phi_\pm(\sqrt{k_1^2+\lambda})}.~~~
\end{eqnarray}
The representation \eqref{A=} clearly follows from \eqref{Atilde=2} once we prove that $\tilde H^\pm_{k_1}(\lambda)$ is the Laplace transform of a continuous function $H^\pm_{k_1}(t)$ satisfying \eqref{H}.
By \eqref{Q=}, we see that the singularities of $\tilde H^\pm_{k_1}(\lambda)$ lie on the horizontal half-line $\{\lambda\in\bb C;\,\Re\lambda \le - k_1^2, \,\Im\lambda=0\}$. The last term on the right side of \eqref{Htilde} is not singular at $\lambda=0$ since the denominator has only a simple zero at that point. Then, by the Laplace inverse formula, setting $\zeta=\lambda+k_1^2$, we find
\begin{equation*}
H^\pm_{k_1}(t) = \frac{\rme^{-k_1^2 t}}{2\pi\rmi}  \int_{a-\rmi\infty}^{a+\rmi\infty}\!\rmd \zeta\, \rme^{\zeta t}\, \tilde H^\pm_{k_1}(\zeta-k_1^2) \qquad \forall\, a>0.
\end{equation*}
Choosing $a=\gamma k_1^2$ and $\zeta= k_1^2 (\gamma + \rmi y)$, we get
\begin{equation*}
\begin{split}
H^\pm_{k_1}(t) =\, & \frac{k_1^2\,d_\pm(k_1)}{\pi^2} \exp\left[-(1-\gamma)\, k_1^2t\right] \\ & \times \Bigg\{\int_{\bb R}\!\rmd y\,\sum_{n=0}^4 \left[\frac{d_\pm(k_1)}{|k_1|\sqrt{\gamma + \rmi y}}\right]^n\left[\phi_\pm\left(|k_1|\sqrt{\gamma + \rmi y}\right) - 1 \right] \rme^{\rmi y t}\,(\gamma + \rmi y)\\ & ~~~~+ \int_{\bb R}\!\rmd y\,\left[\frac{d_\pm(k_1)}{|k_1|\sqrt{\gamma + \rmi y}}\right]^4 \frac{\rme^{\rmi y t}\,(\gamma + \rmi y)}{|k_1|\sqrt{\gamma + \rmi y} - d_\pm(k_1)\, \phi_\pm\left(|k_1|\sqrt{\gamma + \rmi y}\right)} \Bigg\}.
\end{split}
\end{equation*}
For   $0<\gamma<1$, the integrals on the right side are absolutely convergent, uniformly with respect to $k_1\ne 0$. Hence \eqref{H} follows with a suitable $B_\gamma>0$. The lemma is proved.
\qed
\begin{lemma}
\label{lem:4}
There is a constant $C_N>0$ such that, for any $\omega,\tilde\omega\in \Omega_{\alpha,\beta,T}$ satisfying \eqref{p14} we have 
\begin{equation}
\label{N2}
\left| N_{k_1,k_2}[\omega(t)] -N_{k_1,k_2}[\tilde\omega(t)] \right| \, \le \, \frac{C_N\, \rme^{-(1+|k_1|)t/4}}{1+|k_1|^\beta}\, R(\omega,\tilde\omega)\, \|\omega-\tilde\omega\|_{\alpha,\beta,T},
\end{equation}
where 
\begin{equation}
\label{R}
R(\omega,\tilde\omega) :=\max\left\{\|\omega\|_{\alpha,\beta,T}; \|\tilde\omega\|_{\alpha,\beta,T}\right\}.
\end{equation}
\end{lemma}
\proof Observe first   that, by  \eqref{N00}, $N_{0,0}[\omega(t)]=N_{0,0}[\tilde\omega(t)]=0$. From the definition \eqref{p13}, as  $\hat\omega_{0,0}=0$, we get, for any $t\in[0,T]$,
\begin{equation*}
\begin{split}
\big| N_{k_1,k_2}[\omega(t)] & - N_{k_1,k_2}[\tilde\omega(t)] \big| \, \le \, \rme^{-(1+|k_1|)t/4} \, 2R(\omega,\tilde\omega)\, \|\omega-\tilde\omega\|_{\alpha,\beta,T} \\ & \times \, \sum_{h_2\in\bb Z} \frac{\delta_\mathrm{odd}(h_2+k_2)}\pi \,\left|\frac{2h_2}{h_2^2-k_2^2}\right| \, S_{k_1,h_2}, 
\end{split}
\end{equation*}
where
\begin{equation*}
S_{k_1,h_2} = \sum_{\due{j+\ell = (k_1,h_2)}{j,\ell\ne (0,0)}} \frac{\left|j^\perp\cdot \ell\right|}{|j|^{\alpha+2} \, |\ell|^\alpha} \, \frac{1}{(1+|j_1|^\beta)(1+|\ell_1|^\beta)}.
\end{equation*}
Since, for a suitable constant $C_1>0$,
\begin{equation*}
\frac{1}{(1+|j_1|^\beta)(1+|\ell_1|^\beta)} \le  \frac{C_1}{1+|j_1+\ell_1|^\beta},
\end{equation*}
to prove \eqref{N2} it is enough to show that, for some $C_2>0$,
\begin{equation}
\label{mN1}
\Sigma_{q_1,q_2} := \sum_{\due{j+\ell = q}{j,\ell\ne (0,0)}} \frac{\left|j^\perp\cdot \ell\right|}{|j|^{\alpha+2} \, |\ell|^\alpha}  =
\sum_{\due{j\ne (0,0)}{j\ne q}} \frac{\left|q^\perp\cdot j\right|}{|j|^{\alpha+2} \, |q-j|^\alpha} \le \frac{C_2}{|q|^{\alpha-1}},
\end{equation}
with $q=(q_1,q_2)\in\bb Z^2\setminus (0,0)$. We decompose the sum as  $\Sigma_{q_1,q_2} = \Sigma^{(0)}_{q_1,q_2} + \Sigma^{(1)}_{q_1,q_2} + \Sigma^{(2)}_{q_1,q_2}$ with
\begin{equation*}
\Sigma^{(1)}_{q_1,q_2} := \sum_{\due{0<|j| \le 2|q|}{|j-q| > |q|/2}} \frac{\left|q^\perp\cdot j\right|}{|j|^{\alpha+2} \, |q-j|^\alpha}, \qquad
\Sigma^{(2)}_{q_1,q_2} := \sum_{0<|j-q| \le |q|/2}\frac{\left|q^\perp\cdot j\right|}{|j|^{\alpha+2} \, |q-j|^\alpha}.
\end{equation*}
By elementary inequalities, we see that there are suitable positive constants $C_3, C_4$ such that 
\begin{equation*}
\Sigma^{(1)}_{q_1,q_2} \le \frac{4}{|q|^{\alpha-1}} \sum_{0<|j| \le 2|q|}\frac{1}{|j|^{\alpha+2}} \le \frac{C_3}{|q|^{\alpha-1}}, 
\end{equation*}
\begin{equation*}
\Sigma^{(2)}_{q_1,q_2} \le \frac{16}{|q|^{\alpha+1}} \sum_{0<|j-q| \le |q|/2} \frac{1}{|q-j|^{\alpha-1}} \le \frac{C_4}{|q|^{2\alpha-2}}.
\end{equation*}
Finally, regarding $\Sigma^{(0)}_{q_1,q_2}$, since $|j|>2|q|$ implies $|q-j|\ge |j|$,
\begin{eqnarray*}
\Sigma^{(0)}_{q_1,q_2} & = & \sum_{|j|>2|q|} \frac{\left|q^\perp\cdot j\right|}{|j|^{\alpha+2} \, |q-j|^\alpha} \; \le \; \sum_{|j|>2|q|} \frac{|q|}{|j|^{2\alpha+1}} \\ & \le & \frac{1}{|q|^{2\alpha-3}} \sum_{|j|>2|q|} \frac{1}{|j|^3}\; \le \; \frac{C_5}{|q|^{2\alpha-2}},
\end{eqnarray*}
for a suitable $C_5>0$. The lemma is proved.
\qed
\begin{lemma}
\label{lem:5}
There is a constant $C_*>0$ such that if $\omega,\tilde\omega\in \Omega_{\alpha,\beta,T}$ satisfy \eqref{p14} and $f_{\pm,k_1}(t)$,   $\tilde f_{\pm,k_1}(t)$ are the solutions to \eqref{p18} for $ \omega$ and  $\tilde \omega$, respectively,  the following inequalities hold 
\begin{equation}
\label{f2}
\left|f_{\pm,k_1}(t) - \tilde f_{\pm,k_1}(t) \right| \le \frac{C_*\, \rme^{-(1+|k_1)|t/4}}{1+|k_1|^\beta} \left[|k_1|^{2-\alpha} \, \|\delta\omega\|_{\alpha,\beta,0}+  R(\omega,\tilde\omega)\,\|\delta\omega\|_{\alpha,\beta,T}\right],
\end{equation}
where $\delta\omega := \omega-\tilde\omega$ and $R(\omega,\tilde\omega)$ is defined in \eqref{R}.
\end{lemma}
\proof Introducing the notation 
\begin{equation*}
\delta g_{\pm,k_1}(t) = g_{\pm,k_1}[t;\omega] - g_{\pm,k_1}[t;\tilde\omega], \quad \delta g_{\pm,k_1}'(t) = \partial_t g_{\pm,k_1}[t;\omega] - \partial_t g_{\pm,k_1}[t;\tilde\omega],
\end{equation*}
we see, by \eqref{p17} and \eqref{N2}, that,  for some $C_6>0$, we have,
\begin{eqnarray}
\label{Dg}
&& \!\!\!\!\!\!\!\!\!\!\!\!\!\!\!\!\!\!  \big|\delta g_{\pm,k_1}(t) \big| \,\le \, \frac{1}{1+|k_1|^\beta} \sum_{k_2,\pm} \frac{\rme^{-k^2t}}{|k|^{\alpha+2}}\, \|\delta\omega\|_{\alpha,\beta,0}\,\nonumber \\ && ~~~~~~~~~+\, \frac{1}{1+|k_1|^\beta} \sum_{k_2,\pm} \frac{2\, C_N\,R(\omega,\tilde\omega)\rme^{-(1+|k_1|)t/4}}{|k|^4}\, \|\delta\omega\|_{\alpha,\beta,T}  \nonumber \\ && \!\!\!\!\!\!\!\! \le \, \frac{C_6}{1+|k_1|^\beta} \left[\frac{\|\delta\omega\|_{\alpha,\beta,0}}{|k_1|^{\alpha+1}}\,\rme^{-k_1^2t} + \frac{R(\omega,\tilde\omega)\|\delta\omega\|_{\alpha,\beta,T}}{|k_1|^3}\,\rme^{-(1+|k_1|)t/4}\right]. 
\end{eqnarray}
Here we used the simple inequality
\begin{equation}
\label{kk1}
\rme^{-k^2(t-s)} \rme^{-(1+|k_1|)s/4} \le \rme^{-(1+|k_1|)t/4}\, \rme^{-k^2(t-s)/2}\qquad \forall\, k\ne (0,0),
\end{equation}
and the fact that for each $r>1$ there exists a constant $0<c<\infty$ such that $\sum_{k_2,\pm} |k|^{-r} \le  c\, |k_1|^{1-r}$. In a similar way, since
\begin{eqnarray}
\label{dtg}
\partial_t g_{\pm,k_1}[t;\omega] & = & \sum_{k_2,\pm} \rme^{-k^2 t} \omega_{k_1,k_2}(0) + \sum_{k_2,\pm} \frac{N_{k_1,k_2}[\omega(t)]}{k^2} \nonumber \\ &&  -  \sum_{k_2,\pm} \int_0^t\!\rmd s\, \rme^{-k^2(t-s)} \, N_{k_1,k_2}[\omega(s)],
\end{eqnarray}
we get, for some $C_7>0$,
\begin{equation}
\label{Dg'}
\big|\delta g_{\pm,k_1}'(t)| \le \frac{C_7}{1+|k_1|^\beta} \left[\frac{\|\delta\omega\|_{\alpha,\beta,0}}{|k_1|^{\alpha-1}}\,\rme^{-k_1^2t} + \frac{R(\omega,\tilde\omega)\|\delta\omega\|_{\alpha,\beta,T}}{|k_1|} \,\rme^{-(1+|k_1|)t/4}\right].
\end{equation}
On the other hand, by \eqref{A=}, \eqref{G=}, and \eqref{H}, choosing  $\gamma=\frac 18$, we see that 
\begin{eqnarray}
\label{ff}
&& \!\!\! \big|f_{\pm,k_1}(t) - \tilde f_{\pm,k_1}(t) \big| \, \le \, \frac 2\pi d_\pm(k_1) \, \big|\delta g_{\pm,k_1}'(t)\big| \nonumber\\ && ~~~~~ +\, \frac 2\pi d_\pm(k_1) \sum_{n=1}^4 \left(\frac{d_\pm(k_1)}{\Gamma(n/2)}\right)^n \int_0^t \!\rmd s\, \rme^{-k_1^2(t-s)}\, (t-s)^{(n-2)/2} \big|\delta g_{\pm,k_1}'(s)\big|\nonumber \\ && ~~~~~ +\, B_{1/8} \, |k_1|^3 \int_0^t \!\rmd s\, \rme^{-7k_1^2(t-s)/8} \big|\delta g_{\pm,k_1}(s)\big|.
\end{eqnarray}
Estimate \eqref{f2} then follows by plugging \eqref{Dg} and \eqref{Dg'} into \eqref{ff}, recalling \eqref{d=}, and observing that if $k_1\neq 0$ is an  integer the following inequalities hold,
\begin{eqnarray*}
\rme^{-k_1^2(t-s)} \rme^{-(1+|k_1|)s/4} &\le & \rme^{-(1+|k_1|)t/4}\, \rme^{-k_1^2(t-s)/2}, \\ \rme^{-7k_1^2(t-s)/8} \rme^{-(1+|k_1|)s/4} &\le & \rme^{-(1+|k_1|)t/4}\, \rme^{-k_1^2(t-s)/4}.
\end{eqnarray*}
We omit further details. 
\qed
\begin{remark}\rm
\label{rem:1}
Observe that, taking $\tilde\omega=0$, the estimates \eqref{N2} and \eqref{f2} give
\begin{equation}
\label{N1}
\left| N_{k_1,k_2}[\omega(t)] \right| \, \le \, \frac{C_N\, \rme^{-(1+|k_1|)t/4}}{1+|k_1|^\beta}\, \|\omega\|_{\alpha,\beta,T}^2,
\end{equation}
\begin{equation}
\label{f1}
\left|f_{\pm,k_1}(t)\right| \, \le \, \frac{C_*\, \rme^{-(1+|k_1|)t/4}}{1+|k_1|^\beta} \left(|k_1|^{2-\alpha}\,\|\omega\|_{\alpha,\beta,0}+ \|\omega\|_{\alpha,\beta,T}^2 \right).
\end{equation}
\end{remark}

\medskip
 We now define the iteration procedure for the solution of equations \eqref{p16}-\eqref{p18}.  At the first step we set
\begin{equation*}
\omega_{k_1,k_2}^{(0)}(t) := \rme^{-k^2 t}\,\omega_{k_1,k_2}(0), \qquad
g_{\pm,k_1}^{(0)}(t) :=  - \sum_{k_2,\pm} \frac{1}{k^2} \rme^{-k^2 t} \,\omega_{k_1,k_2}(0), 
\end{equation*}
and denote by $f_{\pm,k_1}^{(0)}(t)$ the solution of the Volterra equation
\begin{equation*}
\sum_{k_2,\pm} \frac{1}{k^2}  \int_0^t\!\rmd s\, \rme^{-k^2(t-s)} \, f_{\pm,k_1}^{(0)}(s) = g_{\pm,k_1}^{(0)}(t).
\end{equation*}
We then iterate by setting, for each integer $n\ge 1$,
\begin{equation*}
\omega_{k_1,k_2}^{(n)}(t) := \rme^{-k^2t}\omega_{k_1,k_2}(0) + \int_0^t\!\rmd s\, \rme^{-k^2(t-s)} \, \left[f_{\pm,k_1}^{(n-1)}(s) - N_{k_1,k_2}^{(n-1)}(s)\right],
\end{equation*}
where $N_{k_1,k_2}^{(n)}(t) := N_{k_1,k_2}[\omega^{(n-1)}(t)]$ and $f_{\pm,k_1}^{(n)}(t)$ denotes the solution to 
\begin{equation*}
\sum_{k_2,\pm} \frac{1}{k^2}  \int_0^t\!\rmd s\, \rme^{-k^2(t-s)} \, f_{\pm,k_1}^{(n)}(s) = g_{\pm,k_1}^{(n)}(t),
\end{equation*}
where  $g_{\pm,k_1}^{(n)}(t) := g_{\pm,k_1}[t;\omega^{(n-1)}]$ is given by \eqref{p17}. 
Before proving Theorem \ref{thm:3} we need one more lemma.
\begin{lemma}
\label{lem:5a} 
Under the assumptions above, the following assertions hold.

i) There is some  $T_1=T_1(D_0,\alpha,\beta)>0$ such that
\begin{equation}
\label{D1}
\left\|\omega^{(n)}\right\|_{\alpha,\beta,T} \le D_2 \qquad \forall\, n\ge 0, \qquad \forall\,0\le T \le T_1,
\end{equation} with
\begin{equation}
\label{D1=}
D_2 = 2(1+2C_*)D_0,
\end{equation}
$C_*$ being the constant appearing in Lemma~\ref{lem:5}. 

ii) There is some  $T_0=T_0(D_0,\alpha,\beta)$, $0<T_0\le T_1$, such that \begin{equation}
\label{contra}
\left\|\omega^{(n+1)} - \omega^{(n)} \right\|_{\alpha,\beta,T} < \frac 12  \left\|\omega^{(n)} - \omega^{(n-1)} \right\|_{\alpha,\beta,T}\quad \forall\, n\ge 1, \quad \forall\,0\le T \le T_0.
\end{equation}
\end{lemma} 
\proof By \eqref{idi} we have $\left\|\omega^{(0)}\right\|_{\alpha,\beta,T} \le D_0 <  D_2$. Proceed by induction and assume that \eqref{D1} holds for any $0\le n' <n$ up to some time $T_1$. Observe that, as  $\omega^{(n-1)}_{k_1,k_2}(0) = \omega_{k_1,k_2}(0)$ for all    $n=1,2,\ldots$,  we have,
\begin{equation*}
\left\|\omega^{(n-1)}\right\|_{\alpha,\beta,0}\le D_0, \qquad \left\|\omega^{(n-1)}\right\|_{\alpha,\beta,T} \le D_2.
\end{equation*}
Furthermore we have,
\begin{equation*}
\left|\omega_{k_1,k_2}^{(n)}(t)\right| \le \frac{D_0\,\rme^{-k^2 t}}{|k|^\alpha \left(1+|k_1|^\beta\right)} + \int_0^t\!\rmd s\, \rme^{-k^2(t-s)} \, \left[\left|f_{\pm,k_1}^{(n-1)}(s)\right| + \left|N_{k_1,k_2}^{(n-1)}(s)\right|\right].
\end{equation*}
By applying \eqref{N1} and \eqref{f1} on the right side and using \eqref{kk1}, recalling \eqref{D1=} and that $k\ne (0,0)$, we obtain
\begin{eqnarray*}
\left|\omega_{k_1,k_2}^{(n)}(t)\right| & \le & \frac{\rme^{-(1+|k_1|t)/4}}{|k|^\alpha \left(1+|k_1|^\beta\right)}\bigg[ D_0 + 2\frac{1-e^{-k^2t/2}}{|k|^{2-\alpha}} \\ && ~~~~~\times\left(C_* D_0|k_1|^{2-\alpha} +C_* D_2^2 + C_N D_2^2\right)\bigg] \\ &\le &  \frac{\rme^{-(1+|k_1|t)/4}}{|k|^\alpha \left(1+|k_1|^\beta\right)}\left[ \frac{D_2}2  + \bar C D_2^2\, t^{(2-\alpha)/2}\right],
\end{eqnarray*}
where 
\begin{equation*}
\bar C := 2\,(C_*+C_N)\,\sup_{\xi>0}\,\frac{1-e^{-\xi/2}}{\xi^{(2-\alpha)/2}}.
\end{equation*}
By the previous estimate, we have $\|\omega^{(n)}\|_{\alpha,\beta,T} \le D_2$ if
\begin{equation*}
T \le T_1 := \left(\frac{1}{2\bar C D_2}\right)^{2/(2-\alpha)},
\end{equation*}
whence \eqref{D1} is proved with this choice of $T_1$.

Passing to the proof of \eqref{contra}, let $t\in (0,T_1]$ so that \eqref{D1} holds. We have
\begin{eqnarray*}
\left|\omega_{k_1,k_2}^{(n+1)}(t) - \omega_{k_1,k_2}^{(n)}(t)\right| & \le &  \int_0^t\!\rmd s\, \rme^{-k^2(t-s)} \, \left|f_{\pm,k_1}^{(n)}(s) -f_{\pm,k_1}^{(n-1)}(s)\right|\\ && + \, \int_0^t\!\rmd s\, \rme^{-k^2(t-s)} \, \left|N_{k_1,k_2}^{(n)}(s) - N_{k_1,k_2}^{(n-1)}(s)\right|.
\end{eqnarray*}
By applying \eqref{N2} and \eqref{f2} on the right side, observing that 
\begin{equation*}
\left\|\omega^{(n)}-\omega^{(n-1)}\right\|_{\alpha,\beta, 0} = 0, \qquad
R(\omega^{(n)},\omega^{(n-1)}) \le D_2,
\end{equation*}
and using \eqref{kk1} as before, we get
\begin{equation*}
\left|\omega_{k_1,k_2}^{(n+1)}(t) - \omega_{k_1,k_2}^{(n)}(t)\right| \le \frac{\rme^{-(1+|k_1|t)/4} \bar CD_2\, t^{(2-\alpha)/2}}{|k|^\alpha \left(1+|k_1|^\beta\right)}  \left\|\omega^{(n)}-\omega^{(n-1)}\right\|_{\alpha,\beta, T},
\end{equation*}
whence \eqref{contra} is proved with $T_0=T_1$.
\qed

\medskip
\noindent{\bf Proof of Theorem \ref{thm:3}.} By Lemma~\ref{lem:5a}, we  have that $\{\omega^{(n)}\}$ is a Cauchy sequence in $\Omega_{\alpha,\beta,T}$ for $0< T<T_0$ satisfying \eqref{D1}. This proves the  existence and uniqueness of the solution in $\Omega_{\alpha,\beta,T}$ for $0<T<T_0$ and that \eqref{tde1} holds with $D_2$ given by \eqref{D1=}.

\smallskip
The proof of the existence of a global solution for small initial data  follows along the same lines with an obvious modification: in proving \eqref{D1} and \eqref{contra}, the small parameter is $D_2$ and the time power $t^{(2-\alpha)/2}$ should not  be extracted. We omit the details.
\qed
\begin{remark}\rm
\label{rem:2}
The decay factor $\rme^{-(1+|k_1|)t/4}$ could be omitted with minor chan\-ges in the proof of Theorem~\ref{thm:3}. In particular, if we only assume that $\omega_{k_1,k_2}(0)$, with $\omega_{0,0}(0)=0$,  satisfies the conditions  \eqref{p14} and
\begin{equation*}
\left|\omega_{k_1,k_2}(0)\right| \le \frac{D_0}{|k|^{\alpha}} \qquad \forall\, k_1\in\bb Z,\, k_2\ge 0,\, k\ne (0,0),
\end{equation*}
with $1 < \alpha < 2$, then there exist a time $T_0=T_0(D_0,\alpha)$ and a unique solution $\{\omega_{k_1,k_2}(t);\,  k_1\in\bb Z,\, k_2\ge 0\}$ to equations \eqref{p11} and \eqref{p14} such that $\|\omega\|_{\alpha,T_0} <\infty$, where
\begin{equation}
\label{norma}
\|\omega\|_{\alpha,t} := \sup_{s\in [0, t]}\, \sup_{k_1\in\bb Z} \, \sup_{k_2\ge 0}\,\left|\omega_{k_1,k_2}(s)\right| \, |k|^{\alpha}.
\end{equation} 
\end{remark}

\section{Proofs of Theorem \ref{thm:2} and Corollary~\ref{cor:1}}
\label{sec:4}

The proof of Theorem~\ref{thm:2} will be achieved by first extending the  solution considered in Remark~\ref{rem:2} to any positive time. The estimates required  for such an extension are in fact well known properties of the weak solutions to the NS flow. Therefore, we first prove that the  local solutions to our system actually coincide with weak solutions of the NS system  \eqref{p1}-\eqref{p2}. 
\begin{lemma}
\label{lem:6}
Let $\{\omega_{k_1,k_2}(t);\,  k_1\in\bb Z,\, k_2\ge 0\}$, $t\in [0,T]$, be a solution to equations \eqref{p11} and \eqref{p14} such that $\|\omega\|_{\alpha,T} <\infty$, $1 < \alpha < 2$, and let
\begin{equation*}
\omega(x,t) := \sum_{k\in\bb Z^2} \hat \omega_{k_1,k_2}(t) \, \rme^{\rmi k\cdot x}.
\end{equation*}
Then, the velocity field $u(x,t) := \nabla^\perp\Delta_N^{-1}\omega(x,t)$ coincides, for $t\in [0,T]$, with a weak solution to the NS system \eqref{p1}-\eqref{p2}. 
\end{lemma}
\proof By the expansion \eqref{nn}, we have 
\begin{equation*}
u(x,t) = \sum_{\due{k\in\bb Z^2}{k\ne (0,0)}} u_{k_1,k_2}(t) \, \rme^{\rmi k\cdot x}, \qquad u_{k_1,k_2}(t) := -\rmi k^\perp \, \frac{\hat\omega_{k_1,k_2}(t)}{k^2},
\end{equation*}
so that in terms of the norm \eqref{norma} we get the estimate
\begin{equation*}
\sup_{t\in [0,T]} \, \left| u_{k_1,k_2}(t)\right| \le\|\omega\|_{\alpha,T} \, \frac{1}{|k|^{\alpha+1}}.
\end{equation*}
Therefore $u\in L^2([0,T];V)$,  where $V$ is the space of solenoidal vector fields in $H_0^1(\mc C)^2$. For the assertion, it remains to be proved \cite{L,T1,T2} that, for any solenoidal $C^\infty$ vector field $\Phi$ of compact support in $\mc C$,  we have,
\begin{equation*}
\frac{\rmd}{\rmd t} \int_{\mc C}\!\rmd x\, \Phi(x) \cdot u(x,t) =
\int_{\mc C}\!\rmd x\, \Delta\Phi(x) \cdot u(x,t) - \int_{\mc C}\!\rmd x\, \Phi(x)\cdot\big[u(x,t)\cdot\nabla \big] u(x,t).
\end{equation*}
Since $\Phi = \nabla^\perp \phi$ for some $C^\infty$ function $\phi$ with derivatives of compact support in $\mc C$, by Green's formula we get,
\begin{equation*}
\int_{\mc C}\!\rmd x\, \Phi \cdot u = - \int_{\mc C}\!\rmd x\, \phi\, \omega, \quad \int_{\mc C}\!\rmd x\, \Delta\Phi \cdot u = - \int_{\mc C}\!\rmd x\, \Delta\phi\, \omega.
\end{equation*}
Using the identity $u\,{\rm rot}\, u = \frac 12 \nabla^\perp u^2 - (u\cdot\nabla)u^\perp$,  we also find
\begin{equation*}
  \int_{\mc C}\!\rmd x\, \Phi \cdot [u\cdot\nabla] u = -  \int_{\mc C}\!\rmd x\, \nabla\phi\cdot [u\cdot\nabla] u^\perp = - \int_{\mc C}\!\rmd x\, \nabla\phi \cdot u\,\omega .
\end{equation*}
Therefore, we have to prove that, for any $C^\infty$ function $\phi$ with derivatives of compact support in $\mc C$, 
\begin{equation}
\label{wf}
\frac{\rmd}{\rmd t} \int_{\mc C}\!\rmd x\, \phi(x) \, \omega(x,t) =
\int_{\mc C}\!\rmd x\, \Delta\phi(x) \, \omega(x,t) - \int_{\mc C}\!\rmd x\, \nabla\phi(x) \cdot u(x,t)\,\omega(x,t) .
\end{equation}
Going back to equations  \eqref{p11} we find
\begin{eqnarray}
\label{wf1}
\frac{\rmd}{\rmd t} \int_{\mc C}\!\rmd x\, \phi(x) \, \omega(x,t) =  2\pi^2 \sum_{k\in\bb Z^2} \hat\phi_{-k_1,-k_2} \, \frac{\rmd \hat\omega_{k_1,k_2}(t)}{\rmd t}  = \int_{\mc C}\!\rmd x\, \Delta\phi(x) \, \omega(x,t)  \nonumber\\+\, 2\pi^2\sum_{k\in\bb Z^2} \hat\phi_{-k_1,-k_2} f_{\pm,k_1}(t) - 2\pi^2\sum_{k\in\bb Z^2} \hat\phi_{-k_1,-k_2}  \hat N_{k_1,k_2}[\omega(t)]. 
\end{eqnarray}
The first sum on the right side of \eqref{wf1} is zero since $f_{\pm,0}(t) = 0$ and $\sum_{k_2,\pm}\phi_{k_1,k_2}$ $= 0$ for any $k_1\ne 0$ since $\nabla\phi$ has compact support. Regarding the last term on the right side of \eqref{wf1}, since the last sum in \eqref{p13} is an odd function of $h_2$, it is equal to (we omit the explicit dependence on time $t$)
\begin{equation*}
2\pi\rmi \sum_{\due{j,k,\ell}{j,k\ne (0,0)}} \frac{\delta_{\textrm{odd}}(k_2+j_2+\ell_2)}{k_2+j_2+\ell_2} \, \delta_0(k_1+j_1+\ell_1)\, \frac{j^\perp\cdot \ell}{j^2} \, \hat\phi_{k_1,k_2} \, \hat\omega_{j_1,j_2} \, \hat\omega_{\ell_1,\ell_2}. 
\end{equation*}
On the other hand, in terms of the Fourier components, the last term on the right side of \eqref{wf} reads,
\begin{equation*}
-2\pi\rmi \sum_{\due{j,k,\ell}{j,k\ne (0,0)}} \frac{\delta_{\textrm{odd}}(k_2+j_2+\ell_2)}{k_2+j_2+\ell_2} \, \delta_0(k_1+j_1+\ell_1)\, \frac{j^\perp\cdot k}{j^2} \, \hat\phi_{k_1,k_2} \, \hat\omega_{j_1,j_2} \, \hat\omega_{\ell_1,\ell_2}. 
\end{equation*}
Therefore, their difference is
\begin{equation*}
2\pi\rmi \sum_{\due{j,k,\ell}{j,k\ne (0,0)}} \frac{\delta_{\textrm{odd}}(k_2+j_2+\ell_2)}{k_2+j_2+\ell_2} \, \delta_0(k_1+j_1+\ell_1)\, \frac{j^\perp\cdot (k+\ell)}{j^2} \, \hat\phi_{k_1,k_2} \, \hat\omega_{j_1,j_2} \, \hat\omega_{\ell_1,\ell_2}. 
\end{equation*}
Substituting $j^\perp\cdot (k+\ell) = j^\perp\cdot (k+j+\ell) = j_1(k_2+j_2+\ell_2)$, we conclude that the above sum vanishes because of conditions \eqref{p14}. The lemma is proved. 
\qed
\begin{lemma}
\label{lem:7}
Let $\{\omega_{k_1,k_2}(t);\,  k_1\in\bb Z,\, k_2\ge 0\}$, $t\in [0,T]$, be any solution to equations \eqref{p11} and \eqref{p14} such that $\|\omega\|_{\alpha,T} <\infty$ and set
\begin{equation}
\label{ue}
\mc U(t) := \sum_{\due{k\in \bb Z^2}{k\ne (0,0)}} \frac{\big|\hat \omega_{k_1,k_2}(t)\big|^2}{k^2}, \qquad \mc E(t) := \sum_{k\in \bb Z^2} \big|\hat \omega_{k_1,k_2}(t)\big|^2.
\end{equation}
Then
\begin{equation}
\label{en}
\mc U(t) \le \mc U(0) \, \rme^{-t} \qquad \forall\, t\in [0,T], 
\end{equation} 
and there are constants $E_0,\sigma>0$, depending on $\mc E(0)$, $\mc U(0)$, such that 
\begin{equation}
\label{es}
\mc E(t) \le \mc \, E_0 \, \rme^{-\sigma t} \qquad \forall\, t\in [0,T].
\end{equation} 
\end{lemma}
\proof Since
\begin{equation*}
\mc U(t) = \frac{1}{2\pi^2} \int_{\mc C}\!\rmd x\, \big|u(x,t)\big|^2, \qquad \mc E(t) = \frac{1}{2\pi^2}\int_{\mc C}\!\rmd x\, \big|\nabla u(x,t)\big|^2,
\end{equation*}
\eqref{en} and \eqref{es} assert, respectively, the decrease in time of the energy $\mc U$ and that the enstrophy $\mc E$ tends to zero. These are well known properties of the NS flow in the absence of external forces, therefore they apply here by Lemma~\ref{lem:6}. Actually, the energy estimate can be easily proved directly in our setting. In fact, by \eqref{p11}, 
\begin{equation*}
\dot {\mc U}(t) + \mc E(t) = \sum_{\due{k\in \bb Z^2}{k\ne (0,0)}} \frac{\hat\omega_{-k_1,-k_2}(t)}{k^2} f_{\pm,k_1}(t) - \sum_{\due{k\in \bb Z^2}{k\ne (0,0)}} \frac{\hat\omega_{-k_1,-k_2}(t)}{k^2} \hat N_{k_1,k_2}[\omega(t)],
\end{equation*}
where we used $\hat \omega_{-k_1,-k_2}$ is the complex conjugate of $\hat \omega_{k_1,k_1}$ since $\omega$ is real. The first sum on the right side is clearly zero because of \eqref{p14}. Regarding the second one, since the last sum in \eqref{p13} is an odd function of $h_2$, it is equal to (we omit the explicit dependence on time $t$)
\begin{equation*}
- \frac{\rmi}\pi \sum_{\due{j,k,\ell}{j,k\ne (0,0)}} \frac{\delta_{\textrm{odd}}(k_2+j_2+\ell_2)}{k_2+j_2+\ell_2} \, \delta_0(k_1+j_1+\ell_1)\, \frac{j^\perp\cdot \ell}{j^2k^2} \, \hat\omega_{k_1,k_2} \, \hat\omega_{j_1,j_2} \, \hat\omega_{\ell_1,\ell_2}, 
\end{equation*}
which can be rewritten in the symmetric form, 
\begin{equation*}
- \frac{\rmi}{2\pi} \sum_{\due{j,k,\ell}{j,k\ne (0,0)}} \frac{\delta_{\textrm{odd}}(k_2+j_2+\ell_2)}{k_2+j_2+\ell_2} \, \delta_0(k_1+j_1+\ell_1)\, \frac{(k+j)^\perp\cdot \ell}{j^2k^2} \, \hat\omega_{k_1,k_2} \, \hat\omega_{j_1,j_2} \, \hat\omega_{\ell_1,\ell_2}. 
\end{equation*}
Substituting $(k+j)^\perp\cdot \ell = (k+j+\ell)^\perp\cdot \ell = - \ell_1(k_2+j_2+\ell_2)$, we conclude that also the above sum vanishes  again because of \eqref{p14}. In conclusion, $\dot {\mc U}(t) + \mc E(t) = 0$, from which \eqref{en} follows since $\mc E(t) \ge \mc U(t)$. 

Concerning \eqref{es}, we recall that in two dimension, if $u_0\in V$ then the weak solution belongs to $L^\infty([0,T];V)$ for any $T>0$, see \cite[Thm.~III.3.10]{T1}. Furthermore, in the case of non forced NS equation, the enstrophy converges exponentially to $0$ as $t\to +\infty$, see \cite[Thm.~III.3.12]{T1}, whence \eqref{es} holds. The lemma is proved.
\qed
\begin{proposition}
\label{prop:1}
Any local solution $\{\omega_{k_1,k_2}(t);\,  k_1\in\bb Z,\, k_2\ge 0\}$, $t\in [0,T]$, to equations \eqref{p11} and \eqref{p14} such that $\|\omega\|_{\alpha,T} <\infty$, extends uniquely to a global solution. Moreover, if 
\begin{equation}
\label{aat}
|\omega|_{\alpha,t} := \sup_{k_1\in\bb Z} \, \sup_{k_2\ge 0}\,\left|\omega_{k_1,k_2}(t)\right| \, |k|^{\alpha}
\end{equation}
then $|\omega|_{\alpha,t}\to 0$ exponentially fast as $t\to +\infty$.
\end{proposition}
\proof Since $|\omega|_{\alpha,T} <\infty$, by Remark~\ref{rem:2} we can start by taking $\omega_{k_1, k_2}(T)$ as initial data, and obtain existence up to a time $T_1 > T$. Iterating the procedure we get a growing sequence of times $T_{j+1} > T_j$, $j\ge 1$, and if $T^*=\lim_{j\to\infty} T_j $, then  $[0,T^*)$ is the maximal interval of existence of the solution. Our goal is to show that $T^* = +\infty$.  If  $T^*<+\infty$ then clearly 
\begin{equation}\label{abs}
\limsup_{t\uparrow T^*} \|\omega\|_{\alpha,t} = + \infty.
\end{equation}
We shall prove that \eqref{abs} is impossible  if $T^*<+\infty$. By \eqref{p16}, 
\begin{equation}
\label{a<}
\left|\omega_{k_1,k_2}(t)\right| \le \frac{\rme^{-k^2 t}|\omega|_{\alpha,0}}{|k|^\alpha} + \int_0^t\!\rmd s\, \rme^{-k^2(t-s)} \, \big[\left|f_{\pm,k_1}(s)\right| + \left|N_{k_1,k_2}[\omega(s)]\right|\big].
\end{equation}
We need an estimate of the terms on the right side of \eqref{a<}.  
For $\left|N_{k_1,k_2}[\omega(t)]\right|$, we argue in analogy to the proof of Lemma~\ref{lem:4}. We have, by \eqref{p13}, 
\begin{equation}
\label{N3}
\left|N_{k_1,k_2}[\omega(t)]\right| \, \le \, \|\omega\|_{\alpha,t} \, \sum_{h_2\in\bb Z} \frac{\delta_\mathrm{odd}(h_2+k_2)}\pi \,\left|\frac{2h_2}{h_2^2-k_2^2}\right| \, \tilde S_{k_1,h_2}[\omega(t)],
\end{equation}
where
\begin{equation*}
\tilde S_{q_1,q_2}[\omega(t)] := \sum_{\due{j\ne (0,0)}{j\ne q}} \frac{\left|j^\perp\cdot (j-q)\right|}{j^2|j-q|^\alpha} \, |\hat \omega_{j_1,j_2}(t)|, \qquad q=(q_1,q_2)\in\bb Z^2.
\end{equation*}
By  the Cauchy-Schwartz inequality, we have 
\begin{equation}\label{CS}
\tilde S_{q_1,q_2}[\omega(t)] \le \sqrt{\mc E(t)} \, \sqrt{\tilde\Sigma_{q_1,q_2}}, \qquad \tilde\Sigma_{q_1,q_2} := \sum_{\due{j\ne (0,0)}{j\ne q}} \frac{1}{j^2|j-q|^{2\alpha-2}},
\end{equation}
where $\mc E(t)$ is defined in \eqref{ue}. We set  $\tilde \Sigma_{q_1,q_2} = \tilde\Sigma^{(0)}_{q_1,q_2} + \tilde\Sigma^{(1)}_{q_1,q_2} + \tilde \Sigma^{(2)}_{q_1,q_2}$ with
\begin{equation*}
\tilde\Sigma^{(1)}_{q_1,q_2} := \sum_{\due{0<|j| \le 2|q|}{|j-q| > |q|/2}} \frac{1}{j^2|j-q|^{2\alpha-2}}, \qquad \tilde\Sigma^{(2)}_{q_1,q_2} := \sum_{0<|j-q| \le |q|/2}\frac{1}{j^2|j-q|^{2\alpha-2}}.
\end{equation*}
Then, by elementary inequalities we find, for some constant $C_8>0$,
\begin{equation*}
\tilde\Sigma^{(1)}_{q_1,q_2} \le \frac{4}{|q|^{2\alpha-2}} \sum_{0<|j| \le 2|q|}\frac{1}{j^2} \le \frac{C_8\log(\rme + |q|)}{|q|^{2\alpha-2}},
\end{equation*}
\begin{equation*}
\tilde\Sigma^{(2)}_{q_1,q_2} \le \frac{4}{|q|^2} \sum_{0<|j-q| \le |q|/2} \frac{1}{|q-j|^{2\alpha-2}} \le \frac{C_8}{|q|^{2\alpha-2}},
\end{equation*}
\begin{equation*}
\tilde\Sigma^{(0)}_{q_1,q_2} = \sum_{|j|>2|q|} \frac{1}{j^2|j-q|^{2\alpha-2}} \le  \sum_{|j|>2|q|} \frac{1}{|j|^{2\alpha}} \le \frac{C_8}{|q|^{2\alpha-2}}.
\end{equation*}
Therefore  $\tilde S_{q_1,q_2}[\omega(t)] \le 3 C_8 \sqrt{\mc E(t)} \, |q|^{1-\alpha} \sqrt{\log(\rme+|q|)}$, and using inequalities \eqref{es}  and  \eqref{CS} in \eqref{N3}, recalling that $\alpha>1$,   we   get, for a suitable $C_9>0$,
\begin{equation}
\label{N4}
N_{k_1,k_2}[\omega(t)] \, \le \,  C_9 \,\rme^{-\sigma t/2}\, \|\omega\|_{\alpha,t}.
\end{equation}

For $\left|f_{\pm,k_1}(s)\right|$, we recall that $f_{\pm,0}(t)=0$ and  for $k_1\ne 0$ we proceed as in the proof of Lemma~\ref{lem:5}. By \eqref{p17},  \eqref{dtg}, and \eqref{N4}, we have, for some $C_{10}>0$,
\begin{equation*}
\big|g_{\pm,k_1}[t,\omega]\big| \le C_{10} \left[\frac{|\omega|_{\alpha,0}}{|k_1|^{\alpha+1}} \rme^{-k_1^2t} + \frac{ \|\omega\|_{\alpha,t}}{|k_1|^3}  \, \rme^{-\sigma t/2}\right],
\end{equation*} 
\begin{equation*}
\big|\partial_t g_{\pm,k_1}[t,\omega]\big| \le C_{10} \left[\frac{|\omega|_{\alpha,0}}{|k_1|^{\alpha-1}} \rme^{-k_1^2t} + \frac{ \|\omega\|_{\alpha,t}}{|k_1|} \, \rme^{-\sigma t/2} \right].
\end{equation*}
An   upper bound for $\big|f_{\pm,k_1}(t)\big|$ is given by the right side of \eqref{ff} with $\delta g_{\pm,k_1}(t)$, resp.~$\delta g_{\pm,k_1}'(t)$, replaced by $g_{\pm,k_1}[t,\omega]$, resp.~$\partial_t g_{\pm,k_1}[t,\omega]$. From the above estimates, assuming, without loss of generality, that  $\sigma<\frac 12$, after some straightforward calculations, we get, for some constant $C_{11}>0$, the inequality
\begin{equation}
\label{fkt}
\big|f_{\pm,k_1}(t)\big| \le C_{11} \left[\frac{|\omega|_{\alpha,0}}{|k_1|^{\alpha-2}} \rme^{-k_1^2t/2} +  \|\omega\|_{\alpha,t} \,\rme^{-\sigma t/2} \right].
\end{equation} 

\smallskip
Introducing  the bounds \eqref{N4} and \eqref{fkt} in the estimate \eqref{a<} we find,
\begin{eqnarray*}
\left|\omega_{k_1,k_2}(t)\right| & \le & \frac{\rme^{-k^2 t}|\omega|_{\alpha,0}}{|k|^\alpha} + 2C_{11}\,\delta_{k_1\ne 0}\,  \frac{\rme^{-k_1^2t/2}-e^{-k^2t}}{k_1^2+2k_2^2} |k_1|^{2-\alpha} |\omega|_{\alpha,0}\nonumber \\ && +\, (C_9+C_{11})\int_0^t\!\rmd s\, \rme^{-k^2(t-s)} \, \rme^{-\sigma s/2}  \, \|\omega\|_{\alpha,s}.
\end{eqnarray*}
Taking into account the obvious inequalities
\begin{equation*}
\frac{\rme^{-k_1^2t/2}-e^{-k^2t}}{k_1^2+2k_2^2} |k_1|^{2-\alpha} \le \frac{\rme^{-k_1^2t/4}}{|k|^\alpha} \rme^{-k_1^2t/4}(k_1^2 t)^{1-\frac\alpha 2} \frac{1-\rme^{-k^2t}}{(k^2t)^{1-\frac\alpha 2}} \le \frac{C_{12}\,\rme^{-k_1^2t/4}}{|k|^\alpha},
\end{equation*}
where the constant $C_{12}$ is given by
\begin{equation*}
C_{12} := \bigg(\sup_\xi \rme^{-\xi^2/4} |\xi|^{2-\alpha}\bigg)\bigg(\sup_\xi \frac{1-e^{-\xi^2}}{|\xi|^{2-\alpha}}\bigg),
\end{equation*}
we see that there is a positive  constant $C_{13}$ such that
\begin{equation*}
|k|^\alpha\,\left|\omega_{k_1,k_2}(t)\right| \le C_{13}\,\left [  \rme^{- t/4} |\omega|_{\alpha,0}\, + \int_0^t\!\rmd s\, |k|^\alpha\, \rme^{-k^2(t-s)} \, \rme^{-\sigma s/2} \, \|\omega\|_{\alpha,s} \right ] .
\end{equation*}
Therefore, setting  $M := C_{13}\,\sup_\xi e^{-\xi^2/2}|\xi|^\alpha$, we have
\begin{equation*}
|k|^\alpha\,\left|\omega_{k_1,k_2}(t)\right| \le C_{13}\, \rme^{- t/4} |\omega|_{\alpha,0} + M \int_0^t\!\rmd s\, \frac{\rme^{-k^2(t-s)/2}}{(t-s)^{\alpha/2}} \, \rme^{-\sigma s /2} \, \|\omega\|_{\alpha,s}.
\end{equation*}
Taking the supremum over $k\ne (0,0)$, by the definition  \eqref{aat}, we obtain
\begin{equation}
\label{disa}
|\omega|_{\alpha,t} \le C_{13}\, \rme^{- t/4} |\omega|_{\alpha,0} + \int_0^t\!\rmd s\, K(t,s) \, \|\omega\|_{\alpha,s},
\end{equation}
where $K(t,s)$ is a singular integrable   kernel 
\begin{equation}\label{ker}
K(t,s) =  M \, \frac{\rme^{-(t-s)/2}}{(t-s)^{\alpha/2}} \, \rme^{-\sigma s /2}.
\end{equation}
We have to analyze separately small and large times. Let $T_K>0$ be such that $\max_{t\in [0,T_K]} \int_0^{T_{K}}\!\rmd s\, K(t,s) \le  \frac 12$. By \eqref{disa}, for all $0 \le t < \min \{T_{K},T^*\}$,
\begin{equation*}
|\omega|_{\alpha,t} \le C_{13}\, |\omega|_{\alpha,0} + \frac 12\, \|\omega\|_{\alpha,t},
\end{equation*}
so that $\|\omega\|_{\alpha,t} \le 2C_{13}\, |\omega|_{\alpha,0}$ for any  $0\le t < \min\{T_K;T^*\}$. Hence  $T^*>T_K$ and 
\begin{equation}
\label{graK}
\|\omega\|_{\alpha,T_K} \le 2C_{13}\, |\omega|_{\alpha,0}.
\end{equation} 
We now take $t\in [T_K,T^*)$ and choose $\delta < T_K $ such that $ \frac{2M}{2-\alpha} \, \delta^{(2-\alpha)/2}\le \frac 12$.
Therefore,
\begin{eqnarray*}
|\omega|_{\alpha,t} & \le & C_{13}\, \rme^{- t/4} |\omega|_{\alpha,0} + \int_0^{t-\delta}\!\rmd s\, K(t,s) \, \|\omega\|_{\alpha,s} +\int_{t-\delta}^t\!\rmd s\, K(t,s) \, \|\omega\|_{\alpha,s} \\ &\le & C_{13}\,  |\omega|_{\alpha,0} + M\int_0^{t-\delta}\!\rmd s \, \frac{\|\omega\|_{\alpha,s} }{(t-s)^{\alpha/2}}+ M\,\|\omega\|_{\alpha,t}\int_{t-\delta}^t\!\rmd s\,\frac{1}{(t-s)^{\alpha/2}} \\ &\le & C_{13}\,  |\omega|_{\alpha,0} + M\delta^{-\alpha/2} \int_0^t\!\rmd s \, \|\omega\|_{\alpha,s} + \frac 12 \|\omega\|_{\alpha,t}.
\end{eqnarray*}
Since by \eqref{graK}, 
\begin{equation*}
\|\omega\|_{\alpha,t} \le 2C_{13}\,|\omega|_{\alpha,0} + \sup_{s\in[T_K,t]}|\omega|_{\alpha,s} \qquad \forall\, t\in [T_K,T^*),
\end{equation*}
we finally obtain an integral inequality for $\|\omega\|_{\alpha,t}$,
\begin{equation*}
\|\omega\|_{\alpha,t} \le 6C_{13}\, |\omega|_{\alpha,0} + 2M\delta^{-\alpha/2} \int_0^t\!\rmd s \, \|\omega\|_{\alpha,s},
\end{equation*}
which implies, by the Gronwall Lemma, 
\begin{equation}
\label{gra}
\|\omega\|_{\alpha,t} \le 6C_{13}\, |\omega|_{\alpha,0} \exp(2M\delta^{-\alpha/2}\, t).
\end{equation}
In particular $\|\omega\|_{\alpha,t}$ remains bounded as $t\uparrow T^*$ if $T^*<+\infty$, whence $T^*=+\infty$.
 
\smallskip
We are left with the proof that $|\omega|_{\alpha,t}$ converges exponentially   to zero.  For this we use a bootstrap argument.  From \eqref{ker} we get the inequality \begin{equation*}
K(t,s) \le  M \, \rme^{-\sigma t /2}\, \frac{\rme^{-(t-s)/4}}{(t-s)^{\alpha/2}},
\end{equation*}
which, if inserted on the right side of \eqref{disa}, together with the bound \eqref{gra}, gives
\begin{equation}
\label{disa1}
|\omega|_{\alpha,t} \le C_{13}\, |\omega|_{\alpha,0} \left\{ \rme^{- t/4} + 6M_\alpha \exp\left[\left(2M\delta^{-\alpha/2} - \frac \sigma 2\right) \, t\right] \right\},
\end{equation}
where $M_\alpha = M\int_0^\infty\!\rmd s\, e^{-s/4}s^{-\alpha/2}$. If $2M\delta^{-\alpha/2} < \sigma/2$ the argument is complete, otherwise \eqref{disa1} implies the inequality
\begin{equation}
\label{gra1}
\|\omega\|_{\alpha,t} \le C_{13} \,|\omega|_{\alpha,0} \,(1+6M_\alpha) \, \exp\left[\left(2M\delta^{-\alpha/2} - \frac \sigma 2\right) \, t\right],
\end{equation}
in which the exponential factor has become smaller by $\sigma/2$ with respect to the previous inequality   \eqref{gra}. We can now  insert the new inequality  \eqref{gra1},  instead of \eqref{gra}, on the right side of \eqref{disa}, and it is easy to see that although the pre-factor may grow, the exponential factor again becomes smaller by $\sigma/2$.

We can then iterate the procedure $n$ times, where $n$ is the smallest integer such that $n\sigma/ 2 > 2M\delta^{-\alpha/2}$, thus obtaining the desired exponential decay. The proposition is proved. 
\qed

\medskip
\noindent{\bf Proof of Theorem~\ref{thm:2} (conclusion).} We first show that Proposition~\ref{prop:1} implies the first assertion (inequality \eqref{tde}) of Theorem~\ref{thm:2} for $\beta=0$. In fact, as $|\omega|_{\alpha,t}$ is uniformly bounded, the components $\omega_{k_1,k_2}(t)$ satisfy the condition  \eqref{idi} with constants $D_0$ bounded from above, so that, by  Theorem~\ref{thm:3}, the existence time, taking $\omega_{k_1,k_2}(t)$ as initial data,  are, for any $t$, larger than a fixed positive time $\bar T$. This implies that there exist real numbers $D^*, t^*>0$ such that
\begin{equation}
\label{fin1}
\left|\omega_{k_1,k_2}(t)\right| \le \frac{D^*\, \rme^{-(1+|k_1|)\min\{t;t^*\}/4}}{|k|^{\alpha}} \quad \forall\, k_1\in\bb Z,\, k_2\ge 0,\, k\ne (0,0)\quad \forall\, t\ge 0.
\end{equation}
Furthermore, since $|\omega|_{\alpha,t}\to 0$ as $t\to+\infty$, by the last assertion of Theorem~\ref{thm:3}, concerning the existence of global solution for small initial data, we can find a time $T=T(D_0,\alpha)$  such that 
\begin{equation}
\label{fin2}
\left|\omega_{k_1,k_2}(t)\right| \le \frac{\bar D_2\, \rme^{-(1+|k_1|)(t-T)/4}}{|k|^{\alpha}} \quad \forall\, k_1\in\bb Z,\, k_2\ge 0,\, k\ne (0,0)\quad \forall\, t\ge T,
\end{equation}
where $\bar D_2=D_2(|\omega|_{\alpha,T},\alpha,0)$ (see Theorem~\ref{thm:3} for notation). By \eqref{fin1} and \eqref{fin2} the inequality \eqref{tde} with $\beta=0$ is obtained for any $\nu=\nu^*<\frac 14$ and a suitable $D_1=D_1^*>0$.

It is not hard to see that for any $\nu < \nu^*$ such a solution satisfies the inequalities \eqref{tde} with a constant $D_1$ depending on $\beta$, $\nu$ and on the initial data. In fact, the assertion is true for the solution up to the time $T_0$ which is granted by Theorem~\ref{thm:3}. For times $t>T_0$ it is enough to observe that    
\begin{equation*}
|\omega_{k_1,k_2}(t)| \le D_1^* \,\frac{e^{-(1 + |k_1|)\nu^*t}}{|k|^\alpha } \le  C  D_1^*\, \frac{e^{-(1 + |k_1|)\nu t}}{|k|^\alpha (1 + |k_1|^\beta)}, 
\end{equation*}
where $C = \max_{k_1} (1 + |k_1|^\beta) e^{(\nu-\nu^*) (1+ |k_1|) T_0}$.

We now prove the stronger inequality \eqref{idib} by means of an easy bootstrap argument. By \eqref{tde}, since $\nu<\frac 14$, the same reasoning as in the proofs of \eqref{N1}, \eqref{f1} gives
\begin{equation}
\label{N1nn}
\left| N_{k_1,k_2}[\omega(t)] \right| \, \le \, \frac{C_N\, \rme^{-\nu(1+|k_1|)t}}{1+|k_1|^\beta}\, D_1^2,
\end{equation}
\begin{equation}
\label{f1nn}
\left|f_{\pm,k_1}(t)\right| \, \le \, \frac{C_*\, \rme^{-\nu(1+|k_1|)t}}{1+|k_1|^\beta} \left(|k_1|^{2-\alpha}\,D_0 + D_1^2 \right).
\end{equation}
 The integral equation \eqref{p16}, with the inequalities \eqref{N1nn}, \eqref{f1nn}, and \eqref{kk1},  gives, for any $t>0$,
\begin{equation*}
|\omega_{k_1,k_2}(t)|  \le \frac{D_0\, \rme^{-k^2 t}}{|k|^{\alpha}\big(1+|k_1|^\beta\big)} + \frac{2(C_N+C_*)\, \rme^{-\nu(1+|k_1|)t}}{k^2\big(1+|k_1|^\beta\big)} \left(D_0\,|k_1|^{2-\alpha} + D_1^2 \right).
\end{equation*}
It follows that for each $t_0>0 $ there is a constant  $\tilde D_1=\tilde D_1(D_0,t_0,\alpha,\beta)$ for which \eqref{idib} is valid.

Finally, the result of Lemma~\ref{lem:6} extends now to the global solution, therefore the velocity field $u(x,t) = \nabla^\perp\Delta_N^{-1}\omega(x,t)$, with $\omega(x,t)$ as in \eqref{o=}, is solution to the NS system. The theorem is thus proved.
\qed

\medskip
\noindent{\bf Proof of Corollary~\ref{cor:1}.} We recall that the heat kernel on the interval $[0,\pi]$ with Neumann boundary condition reads,
\begin{equation}
\label{gt}
\begin{aligned} g_t(x_2,x_2') & = \frac 1\pi +\frac 2\pi \sum_{k_2\ge 1} \rme^{-k_2^2t} \cos(k_2x_2)\cos(k_2x_2') \\ & = \sum_{n\in\bb Z} \left[p_t(x_2-x_2'+2n\pi) + p_t(x_2+x_2'+2n\pi)\right],\end{aligned}
\end{equation}
where $p_t(z) := (4\pi t)^{-1/2}\exp\{-z^2/(4t)\}$ denotes the heat kernel on the whole line. In particular,
\begin{equation*}
\frac 2\pi \sum_{k_2,\pm} \rme^{-k_2^2t} \cos(k_2x_2) =  g_t(x_2,0) \pm g_t(x_2,\pi).
\end{equation*}
Therefore, by \eqref{p16} the vorticity can be written in the following form,
\begin{equation}
\label{ot=}
\begin{aligned} 
\omega(x,t) & = \sum_{k\in\bb Z^2} \hat \omega_{k_1,k_2}(0)\, \rme^{-k^2 t} \, \rme^{\rmi k\cdot x} \\ &~~ + \sum_{k\in\bb Z^2} \int_0^t\!\rmd s\, \rme^{-k^2 (t-s)} \, \hat N_{k_1,k_2}[\omega(s)] \,\rme^{\rmi k\cdot x} \\ & ~~ + \frac \pi 2\sum_{k_1\in\bb Z} \rme^{\rmi k_1 x_1}\int_0^t\!\rmd s\, \rme^{-k_1^2 (t-s)} g_{t-s}(x_2,0) \left[f_{+,k_1}(s) + f_{-,k_1}(s)\right] \\ & ~~ +\frac \pi 2\sum_{k_1\in\bb Z} \rme^{\rmi k_1 x_1}\int_0^t\!\rmd s\, \rme^{-k_1^2 (t-s)} g_{t-s}(x_2,\pi) \left[f_{+,k_1}(s) - f_{-,k_1}(s)\right].
\end{aligned}
\end{equation}
The first sum on the right clearly defines an infinitely differentiable function on $\bar {\mc C}$ for any $t>0$. For the second sum observe that, by \eqref{mN1} and \cite[Lemma 2.1]{DDS1}, a better estimate than \eqref{N1nn} holds. Namely, for some $\tilde C_N>0$,  
\begin{equation*}
\left|N_{k_1,k_2}[\omega(t)] \right| \, \le \, \frac{\tilde C_N\, \log(\rme + |k_2|) \, \rme^{-\nu(1+|k_1|)t}}{(1+|k_2|^{\alpha-1})(1+|k_1|^\beta)}\,D_1^2.
\end{equation*}
Therefore, for each $t>0$ the second sum on the right side of \eqref{ot=} can be differentiated term by term with respect to the $x_2$-variable, uniformly on $\bar {\mc C}$. It remains to be proved that also the derivatives with respect to   $x_2$ of the last two sums are continuous functions on $\bar{\mc C}$. We consider only the first one, as the analysis of the second is the same. By the representation \eqref{gt} of $g_t$ in terms of $p_t$ and \eqref{f1nn}, the absolute value of each term in the series obtained by replacing $g_{t-s}(x_2,0)$ with $\partial_{x_2}g_{t-s}(x_2,0)$, can be bounded by a constant times 
\begin{equation*}
\frac{\rme^{-\nu(1+|k_1|)t}}{1+|k_1|^\beta} \left(|k_1|^{2-\alpha}\,D_0 + D_1^2 \right) \int_0^t\!\rmd s \,\frac{\rme^{-(x_2-2n\pi)^2/(4s)}}{\sqrt{4\pi s}}\, \frac{|x_2-2n\pi|}{2s},
\end{equation*}
which is easily seen to be summable with respect to $k_1$ and $n$ uniformly for $x_2\in [0,\pi]$ .

\section*{Acknowledgments}

We are indebted to Ya.~G.~Sinai for drawing our attention to the open problems in the theory of the two-dimensonal Navier-Stokes in  bounded domains, and for many useful discussions and comments. This work was partially supported by the GNFM-INDAM and the Italian Ministry of the University.


\end{document}